\documentclass{an}
\usepackage{graphicx}
\usepackage{psfig}
\usepackage{times}
\Volume{01}              
\Year{2002}              
\Month{10}               
\Pagespan{223}{227}      

\def\lesssim{\mathrel{\hbox{\rlap{\hbox{\lower4pt\hbox{$\sim$}}}\hbox{$<$}}}}
\def\gtrsim{\mathrel{\hbox{\rlap{\hbox{\lower4pt\hbox{$\sim$}}}\hbox{$>$}}}}

\begin{document}
\lhead[\thepage]{A.N. Author: Title}
\rhead[Astron. Nachr./AN~{\bf XXX} (200X) X]{\thepage}
\headnote{Astron. Nachr./AN {\bf 32X} (200X) X, XXX--XXX}

\title{Galactic X-ray Survey}

\author{Ken Ebisawa\inst{1,2}\fnmsep\thanks{Powerpoint file of the original presentation is
available at http://isdc.unige.ch/\~{}ebisawa/spain2002.ppt.},
S. Yamauchi\inst{3},  A. Bamba\inst{4}, M. Ueno\inst{4} and S. Senda\inst{4}}
\institute{
INTEGRAL Science Data Centre, Chemin d'\'Ecogia 16, CH-1290 Versoix, Switzerland
\and 
Laboratory for High Energy Astrophysics, NASA/GSFC, Greenbelt, MD 20771, USA
\and 
Faculty of Humanities and Social Sciences, Iwate University,  Ueda 3-18-34, Morioka, Iwate 020-8550, Japan
\and
Department of Physics, Kyoto University, Sakyo-ku, Kyoto 606-8502, Japan
}

\date{Received {\it date will be inserted by the editor}; 
accepted {\it date will be inserted by the editor}} 

\abstract{We review highlights of the results    obtained from recent Galactic X-ray survey observations,
in particular ASCA Galactic center and plane survey and our Chandra deep survey on the $(l,b)
\approx (28\fdg5,0\fdg0)$ region.
Strong hard X-ray diffuse components are observed from Galactic ridge, center and bulge, and they
have  both  thermal  and  non-thermal spectral components. 
Dozens of discrete and extended  sources have been discovered on the Galactic plane,
which  also indicate thermal and/or non-thermal X-ray energy spectra.  They
are often associated with radio sources and are considered to be  SNR candidates.
Most of the hard X-ray point sources in the outer 
part of the Galactic plane are considered to be background AGNs, while fraction  of the
Galactic hard X-ray sources (such as 
quiescent dwarf novae) increases toward the Galactic center.
Most of the soft X-ray sources on the Galactic plane are presumably nearby active stars.
\keywords{Galactic Plane; X-rays; Galactic Diffuse Emission; Supernova Remnants; X-ray sources}
}

\correspondence{ebisawa@obs.unige.ch}

\maketitle

\section{Introduction}

Presence of the hard X-ray ($\gtrsim $ 2 keV) emission from the Galactic plane has been
 recognized since  early 1980's. HEAO1 reported  detection of
the hard X-ray emission from the Galactic ``ridge'', whose
integrated luminosity is $\sim 10^{38}$ erg s$^{-1}$ and  energy spectrum
is softer than that of the  cosmic X-ray background (Worrall et al.\ 1982).  A more precise 
scanning observation
was made with EXOSAT  (Warwick et al.\ 1985), which  manifested global distribution of the
hard X-ray emission over the Milky way.  The extended hard X-ray
emission was observed both  from the Galactic ``ridge'' and ``bulge'' regions.

The Tenma satellite performed several pointing observations on the Galactic
``blank'' fields,  and  detected  omnipresent $\sim$6.7  keV iron K-line emission
(Koyama et al.\ 1986).  This is an evidence that the Galactic
hard X-ray emission is associated with highly ionized plasmas, corresponding to
$kT \approx 6 - 10$ keV.  The Ginga satellite also carried out Galactic scan observations
to study distribution of the diffuse emission.  Thanks to its large effective area, 
Ginga was able to map the distribution of the diffuse emission from iron K-line intensities,
so that  contrast of the diffuse emission relative to
the bright point sources is   significantly enhanced.  In this manner, distribution of the
diffuse emission along the Galactic plane was precisely measured  (Yamauchi and Koyama 1993), and  
concentration of the hot plasma around the Galactic center  (Koyama et al. 1989) was  revealed.

\section{ASCA Galactic Survey}

All the hard X-ray observations before 1993 were carried out with non-imaging instruments,
in which sensitivity is limited by source  confusion.
Consequently, it was hardly possible to resolve dim point sources from 
 diffuse emission, and  to know how much  the point source contribution to the
Galactic diffuse emission is.  
In 1993, ASCA was launched as the first imaging satellite in the hard X-ray band
(Tanaka, Inoue and Holt 1994). 
Above 2 -- 3 keV, the interstellar medium is essentially transparent, so that 
ASCA was for the first time able to search for those hard X-ray sources embedded
deeply in the Galactic plane that  had not been detected by previous soft X-ray
imaging observations.

\begin{figure*}
\centerline{\psfig{file=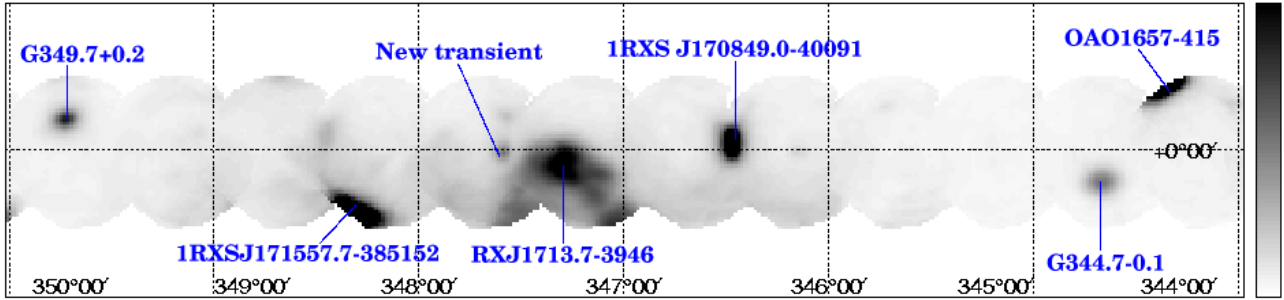,angle=0,width=17.4cm}}
\caption{X-ray image  of a part of  the ASCA Galactic Plane Survey shown with Galactic coordinates.  }
\end{figure*}

In particular, ASCA carried out
systematic survey observations on the Galactic plane and Galactic center regions, and 
acquired  unbiased Galactic hard X-ray imaging data.
The ASCA Galactic plane and center region survey was made to cover the Galactic
inner disk ($|l| < 45^\circ, |b| < 0\fdg4$) and the Galactic
center region ($|l| < 2^\circ, |b| <2^\circ$) with successive pointings of about 10 ksec exposure each.
In addition, there are plenty of non-uniform pointing observations of the Galactic sources or
blank fields.

Most of the ASCA Galactic plane and center survey observation data have been analyzed,
and those results are published.  In the following, some of the highlights  are summarized below.

{\em Point Source Survey:} More than 200 X-ray sources have been resolved in the
 ASCA Galactic  plane and center survey, among which $\sim$60 \% of the sources are unidentifed.
ASCA for the first time made a $\log N$-$\log S$ curve of
Galactic X-ray sources in the 2 -- 10 keV band down to  $\sim 3 \times 10^{-13}$
ergs s$^{-1}$ cm$^{-2}$ (Sugizaki et al.\ 2001).  The $\log N$-$\log S$ curve in this 
flux range was approximated by a power-law with a slope of $\sim 0.8$, which is flatter than
that for the extragalactic sources (Figure 2).
Locations and properties of the
new X-ray sources discovered in the ASCA  surveys are compiled in Sugizaki et al.\ (2001)
and Sakano et al.\ (2002).

{\em Supernova Remnants and Supernova Remnant Candidates:}
ASCA detected X-rays from 30 cataloged radio Supernova Remnants (SNRs) in the
surveyed region, among which 17 SNRs were for the first time detected in X-rays.
In Figure 1, two radio SNRs, G344.7--0.1 and G349.7+0.2 are clearly seen,
whereas they were undetected in ROSAT.  The crescent-like shell feature at $l\sim347^\circ$
is the northwest shell of RX~J1713.7--3946, a SNR discovered with ROSAT.
The ASCA hard X-ray spectrum of  RX~J1713.7--3946 shows non-thermal feature without
emission lines (Koyama et al.\ 1997), which is a reminiscence of SN1006 (Koyama et al.\ 1995).
Later, RX~J1713.7--3946 turned out to be a TeV gamma-ray source (Enomoto et al.\ 2002), just like
SN1006 (Tanimori et al.\ 1998).  These non-thermal X-ray and gamma-ray
emission are considered to be due to synchrotron emission from extremely  energetic 
electrons ($\gtrsim$ TeV), which are presumably accelerated 
by the Fermi mechanism in the expanding SNR shells.  

ASCA also discovered several unidentified extended sources,
some of them show thin thermal X-ray spectra, while others indicate non-thermal
spectra.  These sources may be considered as  X-ray SNR candidates (see Section 5).

{\em  X-ray Pulsars:}
Following  new X-ray pulsars have been discovered in ASCA survey and other pointing observations:
1RXS J170849.0--400910 (P=11s; Sugizaki et al.\ 1997), 
AX J1740.1--2847 (P=729 sec; Sakano et al.\ 2000),
AX J1749.2--725 (P=220 s; Torii et al.\ 1998b),
AX J1820.5--1434 (P=152s; Kinugasa et al.\ 1998),
AX J183220--0840 (P=1549s; Sugizaki et al.\ 2000), 
AX J1841.0--0536 (P=4.7 s; Bamba et al.\ 2001a)
and  AX J1845.0--0300 (P=7s; Torii et al.\ 1998a).
In addition, ASCA discovered several  X-ray sources with flat power-law spectra with large absorption
(characteristics of binary X-ray pulsars), while could not detect coherent pulsations due
to insufficient statistics.  XMM Galactic plane scan survey may be able to  detect coherent
pulsations from these sources.

{\em  Galactic Ridge X-ray Emission:}
ASCA found that the Galactic Ridge energy spectra in 0.5 -- 10 keV are well represented by 
two temperature components (Kaneda et al.\ 1997).  The low temperature
component has a temperature $kT \sim$ 0.8  keV and a low ionization degree. Its surface
brightness is consistent with the SNR origin.  The high temperature component may 
be represented with a temperature  of $kT \sim 7 $ keV in non-ionization equilibrium state
(Kaneda et al.\ 1997).  From the fluctuation analysis of the hard X-ray ridge emission,
it was found that the upper limit of the discrete sources to contribute to the
ridge emission is $\sim 2  \times 10^{31}$ erg s$^{-1}$ (Sugizaki et al.\ 1999).
With that rather benign constraint,  
ASCA was not able to conclude if the Galactic ridge X-ray emission is composed
of numerous discrete sources or truly diffuse emission.

\section{Contribution of the Point Sources}

Chandra's excellent spatial resolution ($\sim0.5''$) revolutionized our X-ray view
of the Milky way.  With $\sim$100 ksec exposure, Chandra is able to detect point sources
down to  a flux of $\sim$3 $\times 10^{-15}$ ergs s$^{-1}$ cm$^{-2}$ in 2 -- 10 keV, that is
about two orders of magnitudes better than ASCA. In Figure 2, we show Chandra $\log N$-$\log S$
curve in 2 -- 10 keV for the Galactic plane ($l=28\fdg5$; Ebisawa et al.\ 2001) and Galactic center 
(Sgr B2 region; Senda 2002). 
Remarkably, the $\log N$-$\log S$ curve on the Galactic plane does not indicate clear  excess of the
Galactic sources compared to the extragalactic  $\log N$-$\log S$ curve.  This indicates
that most of the hard X-ray sources on the Galactic plane are background AGNs at the
Chandra flux limit (Ebisawa et al.\ 2001). The integrated point source X-ray flux 
above $\sim$3 $\times 10^{-15}$ ergs s$^{-1}$ cm$^{-2}$ is only $\sim$ 10 \% of the
total hard X-ray flux in the field of view, which indicates the Galactic ridge
 emission is truly diffuse (Ebisawa et al.\ 2001).  On the other hand, in the 
Galactic center region,  there are  obviously numerous dim point hard X-ray sources
(Figure 2;  Senda 2002), which are presumably mostly quiescent dwarf novae and show
thermal iron K-line emission (Wang, Gotthelf and Lang 2002).  Still, the  integrated point
source flux is $\sim$10 \% of the total flux from the Sgr B2 region, and most of
the hard X-ray emission has truly diffuse origin (Senda 2002).

\begin{figure}
\centerline{\psfig{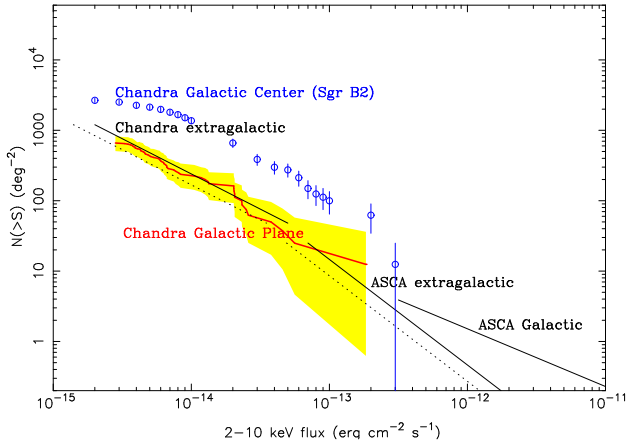}}
\caption{$\log N$-$\log S$ curve in the 2 -- 10 keV band made from Chandra Galactic
plane  (red line with 90 \% error range in yellow; 
Ebisawa et al.\ 2001) and Sgr B2 field (blue; Senda 2002).ASCA Galactic
$\log N$ - $\log S$ curve (Sugizaki et al.\ 2001), and ASCA (Ueda et al.\ 1999) and Chandra 
(Giacconi et al.\ 2001) extragalactic ones are shown together.
}
\end{figure}

\section{Energy Spectra of Galactic Diffuse Emission}

X-ray energy spectra of the diffuse emission from Galactic center, bulge and plane
are very similar in shape (Tanaka 2002).  They have thermal and non-thermal continuum
components, and prominent emission lines from highly ionized iron and other heavy  elements.
The power-law like non-thermal component extends above $\sim$10 keV
(Yamasaki et al.\ 1997; Valinia and Marshall 1998), and smoothly connects to 
the Galactic diffuse gamma-ray emission (Gehrels and Tueller 1993; Valinia, Kinzer and Marshall 2000).

Galactic diffuse X-ray emission is very energetic. Energy density of the
ridge emission,  $\sim$10 eV cm$^{-3}$,  is one or two orders of
magnitude higher than those of  cosmic rays, Galactic magnetic fields or 
any other constituents in the interstellar space.  Energy source of the 
diffuse emission is not elucidated yet, but several attractive theories have been proposed, 
such as interstellar-magnetic reconnection (Tanuma et al.\ 1999), or
interactions of cosmic low-energy electrons (Valinia et al.\ 2000)
or heavy irons (Tanaka 2002) with interstellar medium.

Precise X-ray line study is likely to be a key to resolve origin of the Galactic diffuse emission.
Iron K-line emission feature observed with ASCA has a complex structure, and may not 
be explained by a single equilibrium thermal plasma. 
Kaneda et al.\ (1997) proposes a non-ionization equilibrium plasma model, which
yields a single $\sim$6.6 keV line with a moderate width corresponding to composition
of different ionization states.  Valinia et al.\ (2000) proposes a composite model of the 6.67 keV line
from thermal equilibrium plasma and the 6.4 keV line due to interaction of cosmic-ray
electrons and neutral interstellar matter.  On the other hand, Tanaka (2002)
claims  presence of  the 6.97 keV line which may be attributed to electron capture by the cosmic 
naked iron nuclei. These different models seem to explain the observed ASCA spectra.
Accumulation of more Chandra and XMM data, as well as 
observations by future missions, will eventually 
tell us the definitive answer through precise X-ray spectroscopy.

\section{Discovery of Diffuse and Discrete Sources}
\label{discrete_and_diffuse}
More than a dozen of  diffuse and discrete sources have been discovered with ASCA
and other Galactic surveys.  Most of them are associated with known diffuse radio features,
but some of them are discovered in X-rays for the first time.  

Search for non-thermal X-ray emitting SNRs similar to SN1006 or RX~J1713.7--3946
is very important to study global energy balance of the cosmic ray.
In the ASCA Galactic plane survey,  four such X-ray SNR candidates have been discovered;
G28.6--0.1 (Bamba et al.\ 2001b), G11.0+0.0, G25.5+0.0 and G26.6--0.1 (Bamba et al.\ 2002).
All these sources have power-law slopes of 1.6 to 2.1, without emission lines.
Only G28.6--0.1 is associated with a previously known radio source (Helfand et al.\ 1989; Figure 3).

In contrast to the non-thermal sources, several  diffuse and discrete sources near the Galactic center
such as G0.0--1.3 (Sakano et al.\ 2002) and G0.570--0.018 (Senda, Murakami and Koyama 2002)
show thermal spectra with prominent emission lines.  These thermal sources are likely to be young SNRs
heated by blast waves.

The diffuse X-ray feature in the G28.6--0.1 region has been 
closely studied with ASCA (Bamba et al.\ 2001b) and Chandra (Ebisawa et al.\ 2001; Ueno et al.\ 2002).
The diffuse feature is more clearly seen  in hard X-rays ($>$ 2 keV) than in soft X-rays
(Figure 3). While the extended hard X-ray feature (named AX~J1843.8--0352) shows
a non-thermal spectrum with a slope of $\sim$2.1 (Figure 4, left), 
a blob-like soft X-ray feature (named CXO J18435.1--035828) is found to be embedded 
and associated with the radio emission (Figure 3).  
CXO J18435.1--035828 has a thermal spectrum with prominent emission lines (Figure 4, right).

It is extremely interesting that some discrete and diffuse X-ray sources have non-thermal spectra,
while others show  thermal spectra.  Even there 
is a case like  G28.6--0.1 where thermal and non-thermal components are spatially entangled.
Presumably,  emission mechanism of these thermal and non-thermal components 
from discrete sources  are related  to those of the  global diffuse emission from Galactic center, 
bulge and plane (Section 4).

\section{Origin of the Point Sources}

Characteristics of the point X-ray sources 
detected with Chandra on the Galactic plane at $l\approx28\fdg5$ has been studied,
down to the fluxes $\sim 3 \times 10^{-15}$ erg s$^{-1}$ cm$^{-2}$ (2 --10 keV) or 
$\sim 7 \times 10^{-16}$ erg s$^{-1}$ cm$^{-2}$(0.5 --2 keV)
(Ebisawa et al.\ 2002a).  If the sources are classified with hardness ratio ($HR$) between
0.5--3.0 keV and 3.0--8 keV, the softest sources with $HR \approx -1$ are most numerous, and
the population decreases up to $HR\approx0.5$, then again increases toward  $HR=1$.  This dichotomy
indicates that there are two main populations of the point X-ray sources classified with the spectral
hardness.

Hard X-ray sources are considered to be mostly background AGNs from the argument of the source 
number density (Section 3).  In fact, average hydrogen column densities toward these sources ($\sim 
8 \times 10^{22}$cm$^{-2}$) are
consistent with the value through  the Galactic plane.
However, some of the hard X-ray point sources show flat spectra
(power-law slope $\sim 1$) and iron line feature.  These point sources are candidates of
Galactic hard X-ray sources such as quiescent dwarf novae (e.g., Mukai and Shiokawa 1993).
Soft X-ray sources have low temperature ($\lesssim$ 1 keV) thin thermal spectra, and
low hydrogen column density ($\lesssim  10^{22}$cm$^{-2}$), and some of which show  X-ray flares.
These facts suggest that most of the soft X-ray sources are nearby X-ray active stars.

We have carried out a follow-up observation of this region using the  SOFI infrared
camera at ESO/NTT to identify  these point X-ray sources (Ebisawa et al.\ 2002b).
There are many infrared counterparts of soft X-ray sources brighter than  $K_s \approx 20$ mag,
while very few hard X-ray sources were identified.  This also strongly suggests that 
most soft X-ray sources are nearby stars and hard X-ray sources are background AGNs.

\acknowledgements
Authors are grateful to the following colleagues for supplying 
material presented in this paper:
Kaneda, H., Kinugasa, K., Kokubun, M., Koyama, K., Maeda, Y.,  Matsuzaki, K., 
Mitsuda, K., Murakami, H., Torii, K., Sakano, M. and Sugizaki, M. (ASCA Galactic Survey), 
Paizis, A., Sato, G. (Chandra Galactic plane analysis).

\begin{figure}
\centerline{\psfig{file=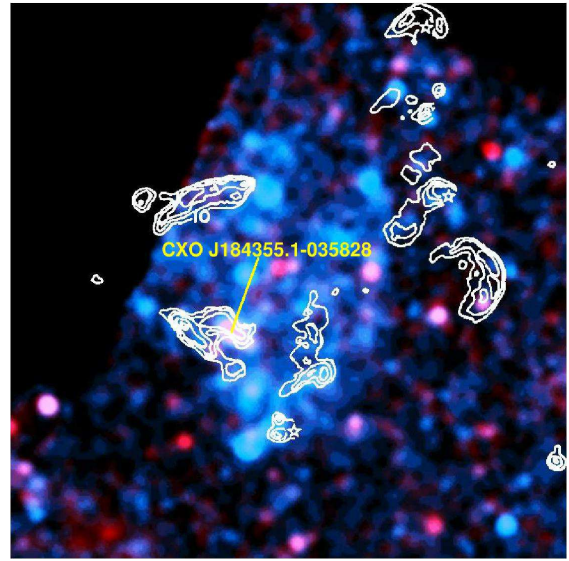,angle=0,width=7.4cm}} 
\caption{A Chandra close-up view of the G28.6--0.1 region overlaid with
radio contours (VLA 20cm; Helfand et al.\ 1989). Red and blue color indicates
soft and hard X-rays respectively.  The extended hard X-ray feature, first
discovered with ASCA,  is named AX~J1843.8--0352 (Bamba et al.\ 2001b).
The extended soft X-ray source CXO J184355.1--035828 is marked  (Ueno et al.\ 2002).
}
\end{figure}

\begin{figure}
\centerline{\psfig{file=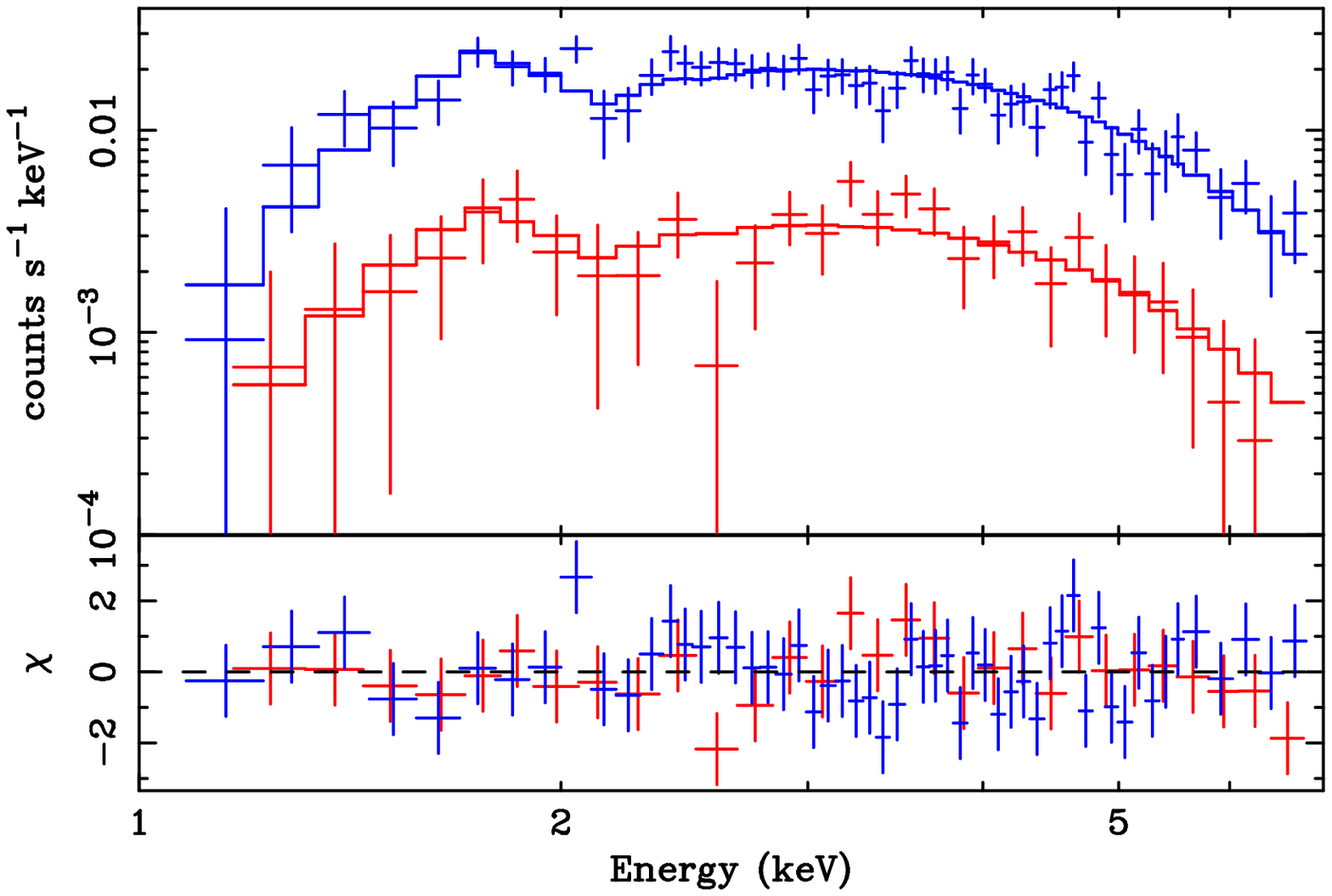,angle=0,width=4.5cm,height=4.9cm}
            \psfig{file=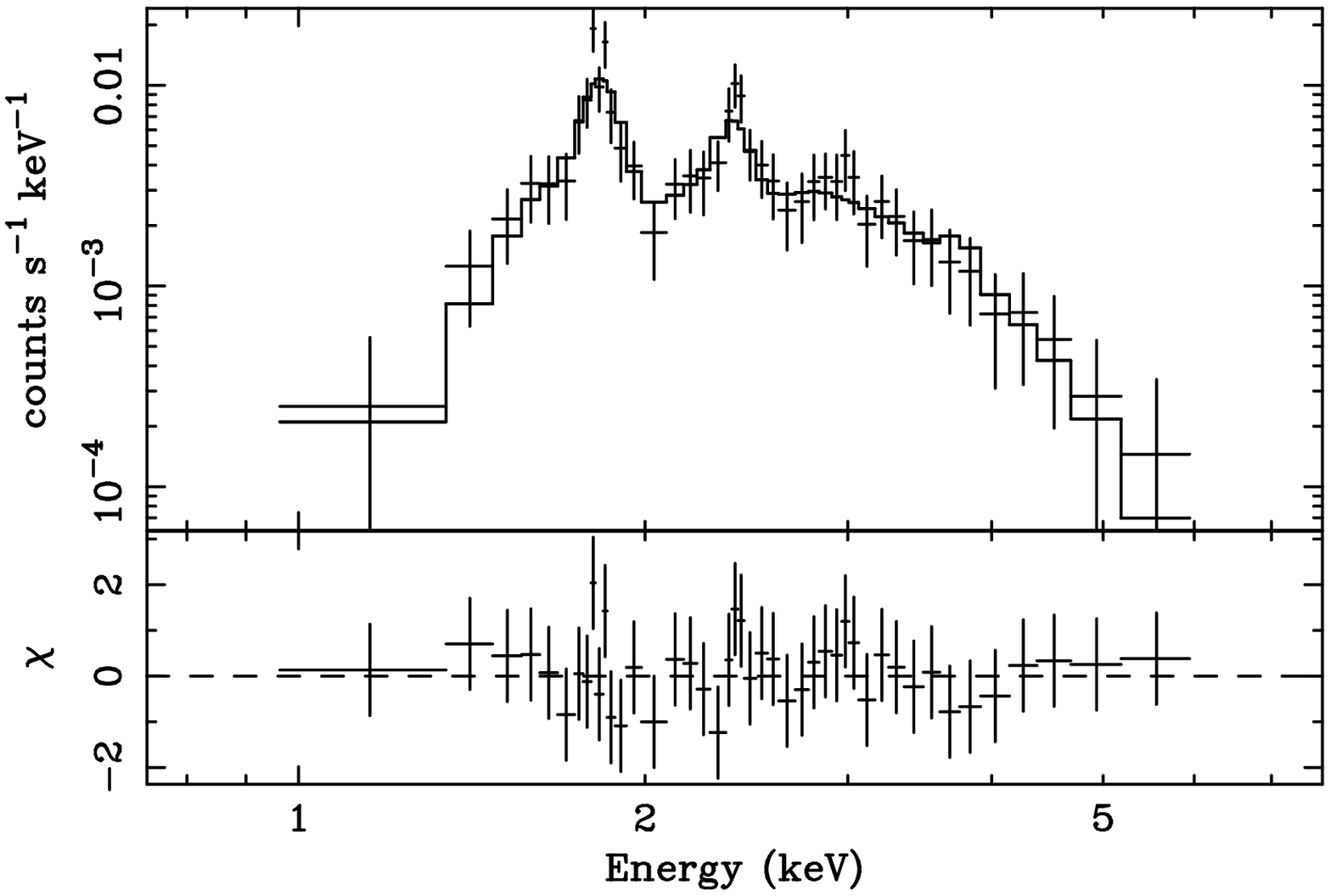,angle=0,width=4.5cm,height=4.9cm}}
\caption{Chandra X-ray energy spectra of AX~J1843.8--0352 (left; 
CXO J184355.1--035828 region is excluded) and
CXO J184355.1--035828 (right) (Ueno et al.\ 2002). 
For AX~J1843.8--0352, two independent observations, one of which only partially
covered the source,  correspond to the two spectra with different normalizations.
}
\end{figure}

\end{document}